\documentclass{PoS}

\usepackage{heppennames2}

\usepackage{lineno}

\graphicspath{{./plots/}}

\title{The fate of the Littlest Higgs with T-parity under 13 TeV LHC data}

\ShortTitle{The Littlest Higgs with T-parity under 13 TeV LHC data}

\author{
  \speaker{J\"urgen Reuter} \\
  DESY, Hamburg, Germany\\
  E-mail: \email{juergen.reuter@desy.de}
}

\author{Daniel Dercks \\
DESY, Hamburg, Germany \\
E-mail: \email{daniel.dercks@desy.de}
}

\author{Gudrid Moortgat-Pick \\
University of Hamburg, Germany \\
E-mail: \email{gudrid.moortgat-pick@desy.de}
}

\author{So Young Shim \\
Konkuk University, Seoul, South Korea \\
E-mail: \email{soyoung.shim@desy.de}
}

\abstract{
  Little Higgs models - which can most easily be thought of as a variant of
  composite Higgs models - explain a light Higgs boson at 125 GeV as an
  pseudo-Nambu-Goldstone boson of a spontaneously broken global symmetry.
  The mechanism of collective symmetry breaking shifts the UV scale of
  these models to the 10 TeV scale and higher. T-parity is introduced as
  a discrete symmetry to remove tree-level constraints on the electroweak
  precision data. Still after run 1 of LHC, electroweak precision observables
  gave stronger constraints than Higgs data and direct searches. We present
  a full recast of all available 13 TeV searches from LHC run 2 to show that
  now direct searches supersede electroweak precision observables. The latest
  exclusion limits on the LHT model will be presented, as well as an outlook
  on the full high-luminosity phase of LHC.
}

\FullConference{39th International Conference on High Energy Physics\\
		4-11 July 2018\\
		Seoul, Korea}

\begin{document}

\section{Introduction}

The Little Higgs mechanism whose most minimal implementation, the
Littlest Higgs model~\cite{ArkaniHamed:2002qy}, gives a solution to the
hierarchy problem: the Higgs is light because it is a
pseudo-Nambu-Golstone boson of some new global symmetry spontaneously
broken at scale $f \sim$ TeV. In order to evade severe constraints
from new strongly interacting sectors at the TeV scale, a collective
symmetry breaking mechanism introduces a quadratically cut-off
sensitive Higgs mass term only at the two-loop level and pushes an
underlying strong sector to the tens of TeV scale. 
Despite this, electroweak precision observables (EWPO,
cf. e.g.~\cite{Kilian:2003xt} still give tight constraints on Little
Higgs models, particu\-larly from the oblique corrections to the vacuum
energies of the weak gauge bosons, such that the symmetry breaking 
scale $f$ is limited to be larger than at least 3 TeV at 95 \%
confidence level. In order to allow for much lower scales, a discrete
symmetry -- TeV or T parity~\cite{Cheng:2003ju} -- has been
introduced. This symmetry relaxes the bound on $f$ by an order of
magnitude, and forces new particles to be pair-produced, and then
undergo cascade decays. The lightest T-odd particle (LTP), the heavy
photon partner $A_H$, is potentially stable. We investigate the
corresponding model, the Littlest Higgs model with T-parity (LHT) and
study the limits from the 8 and 13 TeV runs of the LHC from
direct searches of the new heavy vector bosons, heavy quark and lepton
partners predicted in this model. This study is based
on~\cite{Dercks:2018hgz} as well as on earlier work published
in~\cite{LHM,Reuter:2013iya}. Both due to tight constraints from
searches for Dark Matter as well as the possibility for $T$-breaking
UV completions, we also consider T-parity breaking. 

The model is based on an $SU(5)/SO(5)$ coset space, where the EW gauge
group has been doubled to $\left[SU(2)\times U(1)\right]^2 \to SU(2)_L
\times U(1)_Y$ in order to implement collective symmetry
breaking. T-parity renders the mixing angle 45 degrees. The
implementation of T parity in the fermion sector necessitates the
postulation of vector-like T-odd lepton and quark partners for all
generations as well as T-even and -odd top partners ($T^\pm$). Besides
the symmetry breaking scale $f$, the model depends on the ratio of the
two Yukawa couplings in the top sector, $R := \lambda_1/\lambda_2$,
and the Yukawa couplings of the heavy quark partners, $\kappa_q$ and
heavy lepton partners, $\kappa_\ell$. Furthermore, the model predicts
deviations of order $v^2/f^2$ of the SM charged and neutral current as
well as Higgs couplings from their SM values. This leads to bounds
from EWPO and Higgs measurements that together constrain $f \gtrsim
694$ GeV at 95\% CL~\cite{Reuter:2013iya}. In the section we will
discuss the limits from direct searches.

\section{Recasting of direct searches and indirect constraints}

For the direct searches from Run 1 (8 TeV) and Run 2 (13 TeV) we use
the following notations for the heavy quark partners: $q_H = \left\{
d_H, u_H, s_H, c_H, b_H, t_H \right\}$, the heavy lepton partners:
$\ell_H = \left\{ e_H, \mu_H, \tau_H, \nu_{eH}, \nu_{\mu H}, \nu_{\tau
H} \right\}$, the heavy vectors $V_H = \left\{ W_H, Z_H, A_H \right\}$
as well as the heavy T-even and -odd top partners, $T^\pm$. To compare
with LHC Run 1 and Run 2 data, we studied six different categories of
processes: 1. $p p \rightarrow q_H q_H, q_H \bar{q}_H, \bar{q}_H
\bar{q}_H$, 2. $p p \rightarrow q_H V_H$, 3. $p p \rightarrow \ell_H
\bar{\ell}_H $, 4. $p p \rightarrow V_H V_H$, 5. $p p \rightarrow T^+
\bar{T}^+, T^- \bar{T}^-$, 6. $p p \rightarrow T_+ \bar{q}, \bar{T}_+
q, T_+ W^\pm, \bar{T}_+ W^\pm$. This is pair production of heavy
quarks (1), heavy leptons (3), heavy vectors (4), and heavy top
partners (5) as well as quark-vector (2) and top partner (6)
associated production. We studied a matrix of $2\times 2 \times 3$
different scenarios: first three main scenarios, a so-called
fermion-universality scenario with $\kappa_q = \kappa_\ell = 1.0$, a
light $\ell_H$ scenario with $\kappa_\ell = 0.2$ and a heavy $q_H$
scenario with $\kappa_q = 3.0$. Each of these scenarios was combined
with either light top partners $T^\pm$, i.e. $R = 1.0$, or heavy top
partners inaccessible to the LHC, $R = 0.2$. Furthermore, each of
these scenarios was studied for the T-parity conserving and violating
cases. For the first, the LHC signals are the "classical" SUSY
signals of squark-gluino production or electroweakino/slepton
production, for the case of T-parity violation the $A_H$ decays mainly 
\begin{figure}
\begin{minipage}{0.48\linewidth}\centering
  \includegraphics[width=\textwidth]
                  {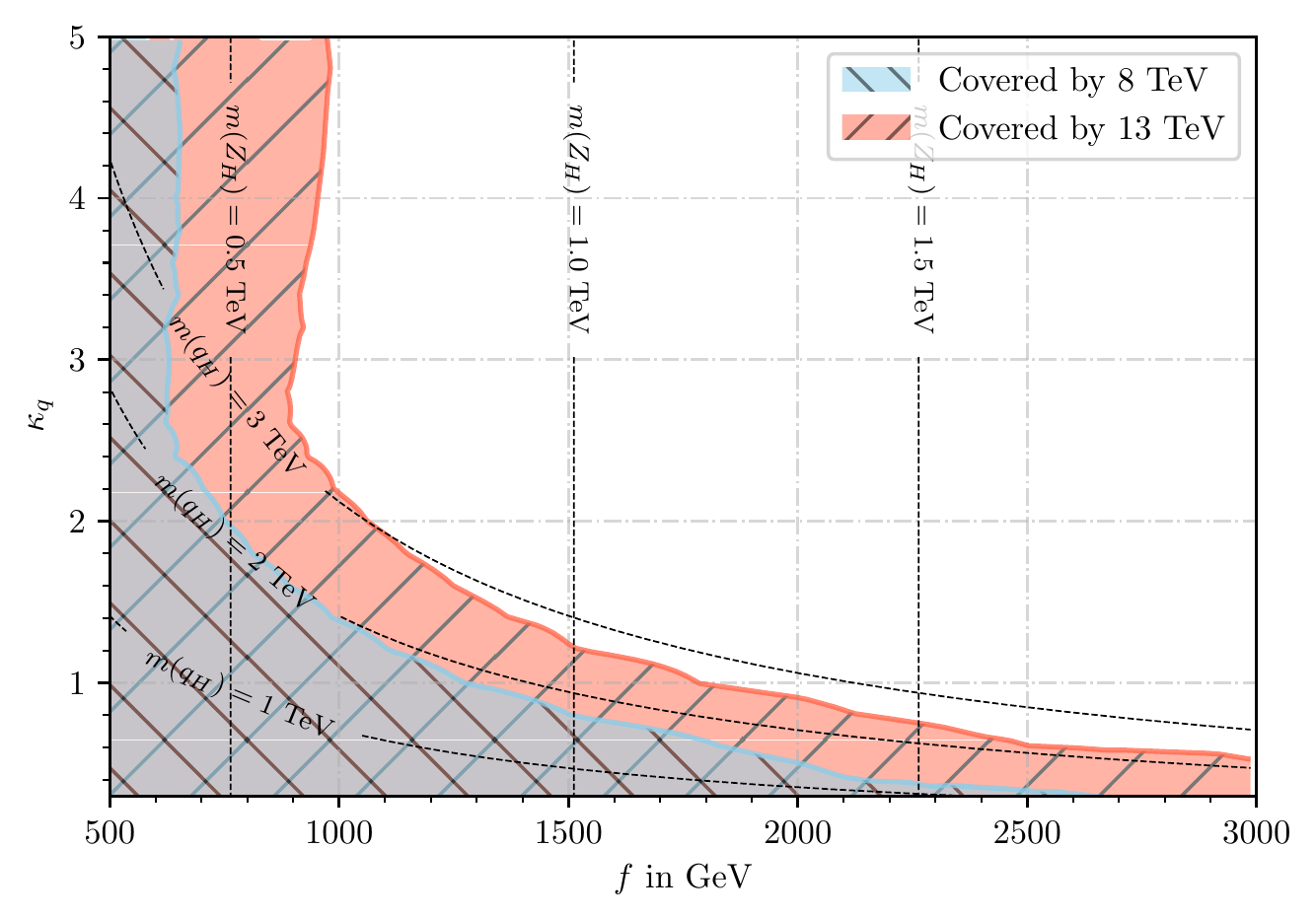}
\end{minipage}
\begin{minipage}{0.51\linewidth}\centering
\includegraphics[width=\textwidth]{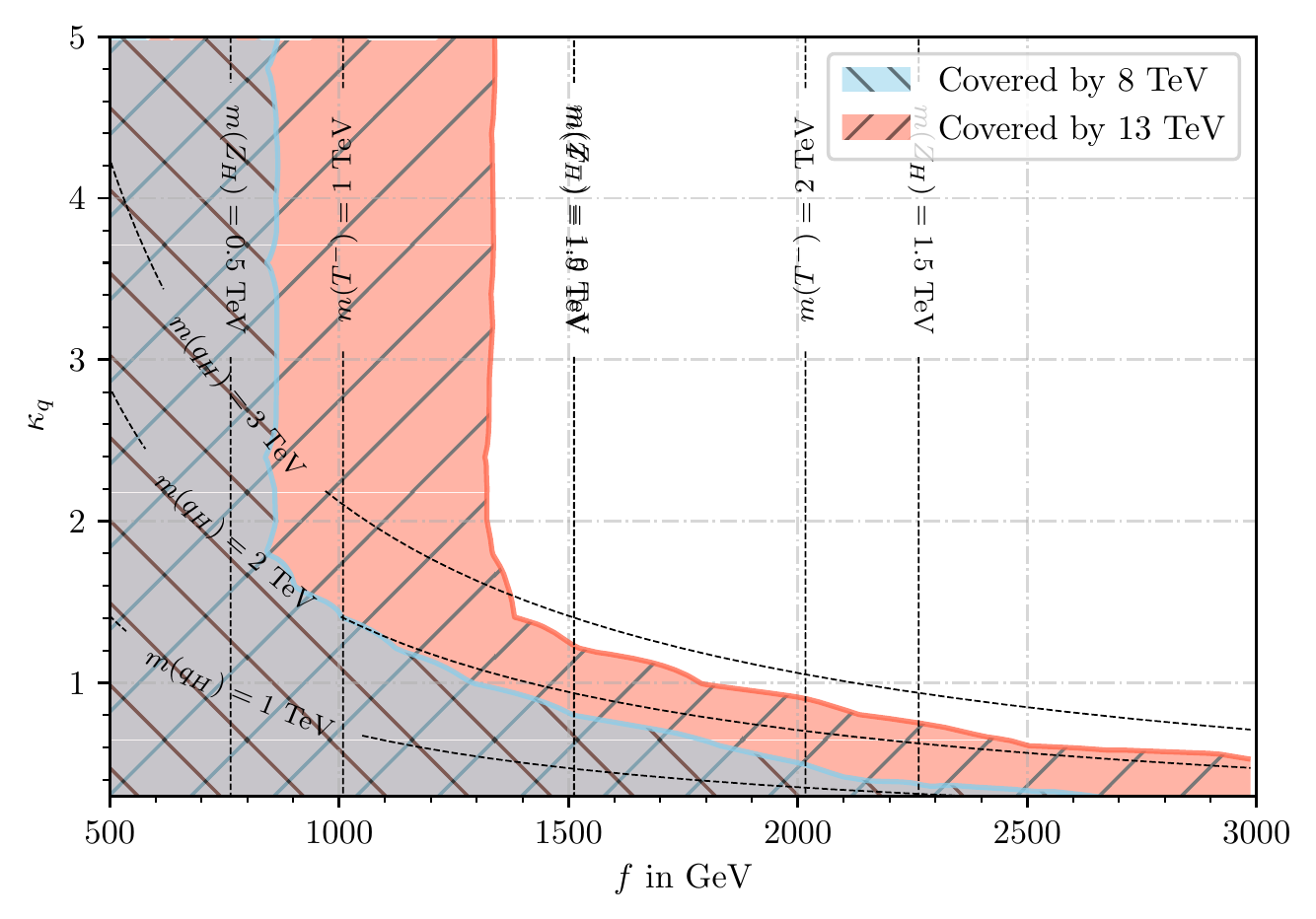}
\end{minipage}
\caption{
  Limits of 8 and 13 TeV data (grey and brown, respectively) for the
  LHT scenario with fermion universality, T-parity conservation and
  light/heavy T-even/T-odd top partners (left/right panel,
  respectively). For more details cf.~\cite{Dercks:2018hgz}
}
\label{fig:tpc}
\end{figure}
via $A_H \to VV$ to SM gauge bosons, which add extra jets and leptons
to the final state selection, but in most cases leaves enough missing
energy by a sufficient number of neutrinos. We used
\texttt{MG5\_aMC@NLO}~\cite{Alwall:2014hca} and
\texttt{WHIZARD}~\cite{WHIZARD} for the event generation,
\texttt{PYTHIA8}~\cite{Sjostrand:2007gs} for parton shower\-ing and
hadronization and \texttt{Delphes}~\cite{deFavereau:2013fsa} for
detector simulation. The recasting of the SUSY searches from ATLAS and
CMS has been done with \texttt{CheckMate}~\cite{CheckMate}. For more
details on the used analyses cf.~\cite{Dercks:2018hgz}.
\begin{figure}
\begin{minipage}{0.48\linewidth}\centering
  \includegraphics[width=\textwidth]
                  {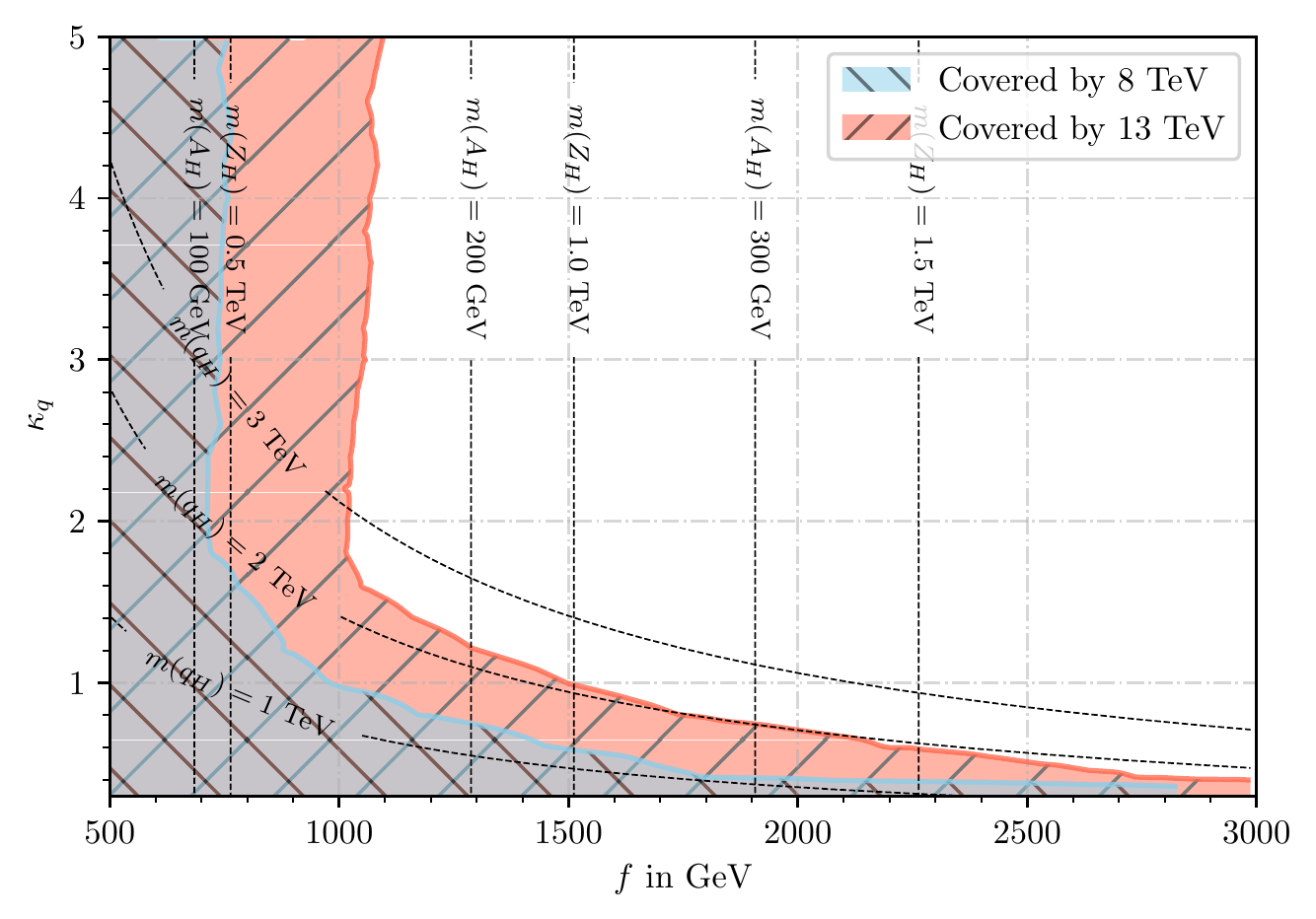}
\end{minipage}
\begin{minipage}{0.51\linewidth}\centering
\includegraphics[width=\textwidth]{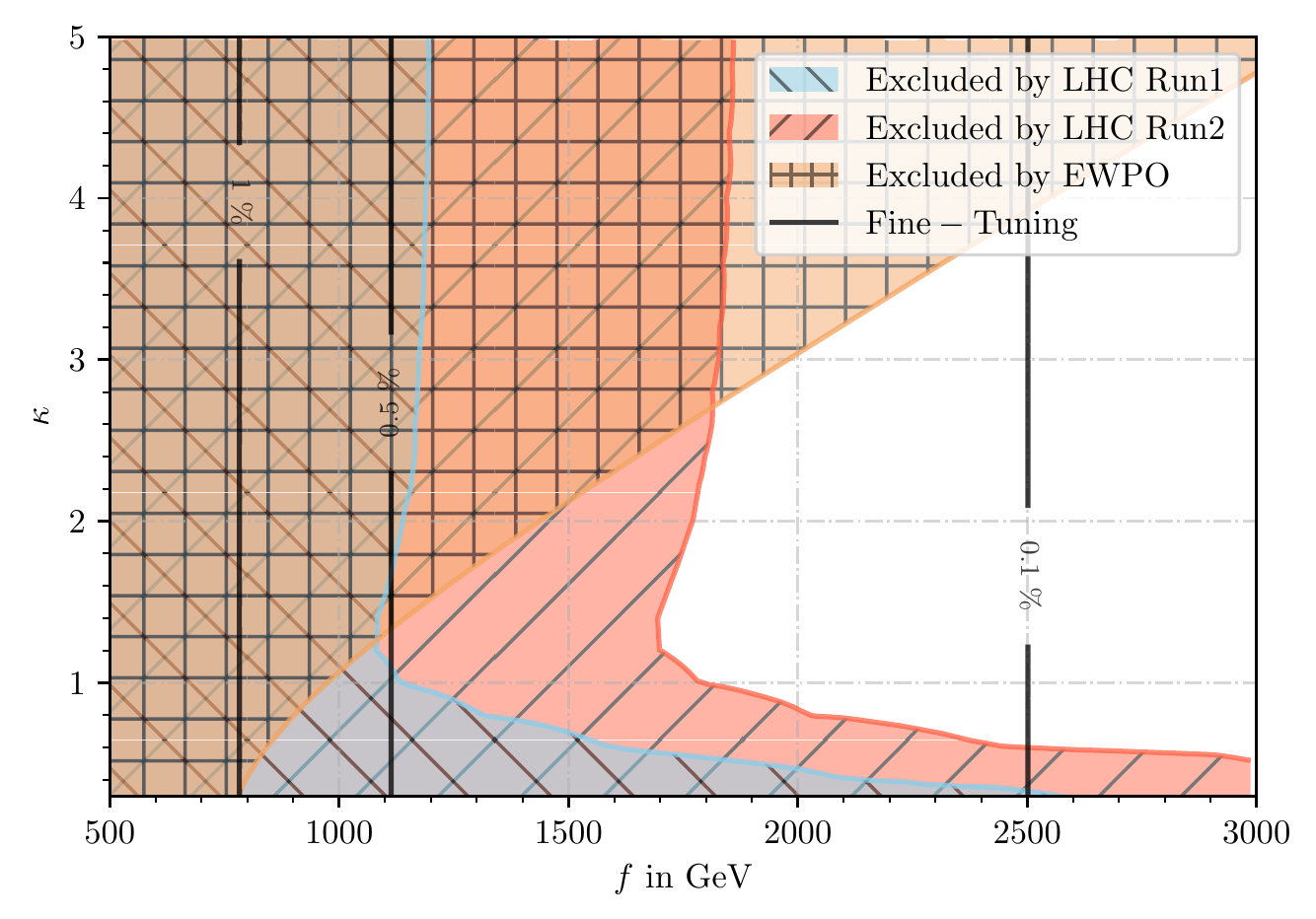}
\end{minipage}
\caption{
  Combination of exclusion for the light $\ell$ and light $T$ scenario
  with T-parity conversation for LHC Run 1 and Run 2 direct
  searches, respectively, as well as from 4-fermion operators. The
  latter are not as model-independent as the direct searches, and
  complementary in exclusion regions compared to the direct searches.
 }
\label{fig:tpv_ewpo}
\end{figure}
For brevity we show only the fermion universality scenario in
Fig.~\ref{fig:tpc} with heavy (left) and light top partners (right).
The most effective analysis is for two jets and missing transverse
energy (MET) searching for $pp \to q_H q_H \to jjA_H A_H + X$. For
high values of $f$, the bounds follow isocontours of $M(q_H) \sim f
\times \kappa_q$. For low $f$, the exclusion is independent of $q_H$
and comes from $V_HV_H$ pair production, cf. Fig~\ref{fig:tpc}
left. If light top partners, $T^\pm$, are accessible, this improves
the low $f$, $\kappa_q$-independent bounds (Fig.~\ref{fig:tpc}
right). 

Considering $T$ parity violation does not grossly change the picture,
as can be seen in the left panel of
Fig.~\ref{fig:tpv_ewpo}. Considering all different cases, the Run 1
limits on the scale $f$ have been increased from ca. 700 GeV to
1300 GeV. Taking into account the full energy of 14 TeV and 3,000
fb${}^{-1}$ of integrated luminosity, these bounds will increase to
1.5-1.8 TeV (95\% CL) bound on the scale $f$.

Besides the direct searches, there are also bounds from four-fermion
operators when integrating out the heavy vectors and heavy
fermions. These are proportional to $\kappa_{q/\ell}^2$. Dijet and
dilepton searches lead to stringent bounds for large values of
$\kappa$ (right panel of Fig.~\ref{fig:tpv_ewpo}). However, these
bounds depend on the details of the UV completion and are not as
reliable as those from direct searches (e.g. small coefficients and
cancellations between operators).

\end{document}